\documentclass[aps,prb, superscriptaddress,showpacs,twocolumn,amssym]{revtex4}
\usepackage{latexsym}
\usepackage{graphicx}
\usepackage{rotating}
\usepackage{longtable}
\usepackage[usenames]{color}
\usepackage[normalem]{ulem}

\input colordvi

\definecolor{orange}{rgb}{1,0.5,0}

\begin{document}

\title{Vacancy-mediated fcc/bcc phase separation in Fe$_{1-x}$Ni$_{x}$ ultrathin films}

\author{T.~O.~Mente\c{s}}
\affiliation{Elettra - Sincrotrone Trieste S.C.p.A., Basovizza, Trieste 34149, Italy}

\author{N.~Stoji\'{c}}
\affiliation{The Abdus Salam International Centre for Theoretical Physics, Strada Costiera 11, 34014 Trieste Italy}
\affiliation{IOM-CNR Democritos, Trieste 34151, Italy}

\author{E.~Vescovo}
\affiliation{National Synchrotron Light Source, Brookhaven National Laboratory, Upton, New York 11973, USA}

\author{J.~M.~Ablett}
\affiliation{Synchrotron Soleil, L'orme des Merisiers, Saint-Aubin, BP 48, Gif-sur-Yvette F-91192, France}

\author{M.~A.~Ni\~{n}o}
\affiliation{IMDEA Nanoscience, Campus Universitario de Cantoblanco, Madrid 28049, Spain}

\author{A.~Locatelli}
\affiliation{Elettra - Sincrotrone Trieste S.C.p.A., Basovizza, Trieste 34149, Italy}

\date{\today}

\begin{abstract}
The phase separation occurring in Fe-Ni thin films near the Invar composition is studied by using 
high resolution spectromicroscopy techniques
and density functional theory calculations.
Annealed at temperatures around $300^{\circ}$C, Fe$_{0.70}$Ni$_{0.30}$ films on W(110) break into micron-sized 
bcc and fcc domains with compositions in agreement with the bulk Fe-Ni phase diagram. 
Ni is found to be the diffusing species in forming the chemical heterogeneity.
The experimentally-determined energy barrier of 1.59~$\pm$~0.09~eV is identified 
as the vacancy formation energy via density functional theory calculations.
Thus, the principal role of the surface in the phase separation
process is attributed to vacancy creation without interstitials.
\end{abstract}


\maketitle

\section{Introduction}

Fe, and to a lesser extent Ni, are the most abundant elements in the Earth's core and mantle.~\cite{donsun1995} 
Their alloys and minerals are important in several disciplines, including materials science and engineering in particular.
Fe-Ni alloys show a rich structural phase diagram as a function of composition and temperature 
and a pronounced magneto-structural coupling.
The latter is most evident in the properties of the Invar alloy (at near 35~at.\%\ Ni), which has a near zero thermal
expansion coefficient due to magnetic interactions.~\cite{schilf1999} 
Another example of the interplay between structure and magnetism
is the divergence of the magnetic spin reorientation
transition thickness in Fe$_{1-x}$Ni$_{x}$ thin films on W(110) with $x$ approaching 0.3.~\cite{lee2006}
Interestingly, the phase diagram~\cite{reuter1989} shows a miscibility gap at just around the same composition range,
in which the alloy has a tendency to separate into fcc and bcc phases with different compositions.
Therefore, understanding the phase decomposition behaviour of Fe-Ni alloy films on W(110) is important
in order to clarify the structural factors contributing to the peculiar magnetic spin reorientation transition (SRT) of this system.

In general, it has been difficult to experimentally observe the structural phase separation in bulk Fe$_{1-x}$Ni$_{x}$ 
near $x\sim0.3$ into spatially distinct regions of different composition. 
Several months of aging at temperatures in the range $300-400^{\circ}$C
results in a heterogeneity at a length scale of about 10~nm.~\cite{zhang1994} Samples collected from meteorites
with a natural treatment at high temperatures and astronomical times show the same phenomenon at the macroscopic
scale.~\cite{bryson2015,kotsugi2010} To speed up the phase separation process,
vacancies were introduced via ion bombardment in order to enhance diffusion 
and rearrangement in the bulk,~\cite{gallas1991, garner1993}
with direct consequences for the magnetic properties.~\cite{chimi2007}

Even in the absence of such conditions that favor phase separation,
the presence of chemical inhomogeneities in the Invar alloy after annealing and a subsequent increase in the
Curie temperature have been reported.~\cite{gallas1991b} 
Similarly, the in-plane to in-plane SRT in Fe$_{1-x}$Ni$_{x}$ thin films on W(110) has been previously 
discussed in the context of a homogeneous strain within the Fe lattice induced by randomly distributed Ni atoms,~\cite{lee2006}
which does not give any explanation for the observed divergence of the reorientation thickness at $x\sim0.3$.
Along these lines, understanding the formation and energetics of heterogeneities in Fe-Ni alloys and alloy films is
a prerequisite in discussing the physics involved both in the Invar effect and in the Fe-Ni/W(110) SRT.

At the limit of ultrathin films, the presence and the nature of phase separation was addressed in a recent work 
on a Fe$_{0.70}$Ni$_{0.30}$ monolayer on W(110).~\cite{mentes2015} 
It was shown that the alloy monolayer decomposes into bcc (pseudomorphic) and
fcc-like (hexagonal) monolayer regions. The Ni content in the bcc and fcc-like regions were measured to be $x=0.15$ and $x=0.42$,
respectively. These compositions are similar to the bulk phase boundaries of Fe-Ni, 
though somewhat shifted towards higher Fe content. 
Nevertheless, the observation of the fcc-like hexagonal layer on the bcc(110) substrate even at the 
monolayer limit is important given that the SRT in Fe films on W(110) is reported to originate 
at the interface layer.~\cite{slezak2010}

In this work, we focus on the fcc-bcc coexistence in thicker Fe$_{1-x}$Ni$_{x}$ ($x\sim0.3$)
films on W(110) at a thickness range closer to the spin reorientation transition.
We show that the phase separation occurs upon annealing a few-nanometers thick alloy film 
to about 300$^{\circ}$C, and the micron-sized
structural domains feature different compositions in agreement with the 
boundaries of the metastability region in the bulk phase diagram.
The observed crystal restructuring requires defect formation (vacancy or interstitial), whereas the chemical 
heterogeneity calls for the diffusion of the alloy species. The high interstitial formation energy above 3~eV 
for bcc Fe~\cite{FuWilOrd04} makes it unlikely at the temperatures considered, which 
leaves vacancy formation as the more likely driving factor of the observed restructuring. 
Based on this, we focus our DFT calculations on the vacancy formation energy in bcc Fe$_{2/3}$Ni$_{1/3}$.
It will be demonstrated that the experimentally-determined
energy barrier of 1.59$\pm$0.09~eV associated to the phase separation kinetics
matches well with the calculations on the vacancy formation.

In the following sections, we begin by describing the experimental details and theoretical methods.
We then present the experimental results on the phase separation in Fe$_{0.70}$Ni$_{0.30}$ films
at high temperature and the chemical and structural characterization of the resulting heterogeneous surface. 
The results of our DFT calculations on the vacancy formation energy and surface diffusion barrier
are reported in section IV, followed by a discussion of the energetics involved in the phase separation process.

\section{Methods}

\subsection{Experimental setup}

The experiments were carried out with the Spectroscopic PhotoEmission and Low-Energy Electron Microscope (SPELEEM)
at the Nanospectroscopy beamline (Elettra Sincrotrone Trieste).~\cite{locatelli2006,mentes2014} The instrument combines
Low-Energy Electron Microscopy (LEEM) with energy-filtered X-ray PhotoEmission Electron Microscopy (XPEEM). 
The former provides structural information,~\cite{altman2010} whereas the latter is used mainly to obtain 
chemical and magnetic distributions.~\cite{locatelli2008} The spatial resolution in the case of LEEM is better than 10~nm, 
whereas in XPEEM the resolution is limited to slightly below 30~nm. In addition to real-space imaging, the instrument can be used to
acquire $\mu$-spot Low Energy Electron Diffraction ($\mu$-LEED) data. 
The transfer width in the diffraction mode is 13.4~nm.
The pressure in the microscope chamber was below $2\times10^{-10}$~mbar during growth and measurements.

The current study took advantage of the fast acquisition rates in LEEM to follow the
growth and phase separation of the FeNi alloy films on W(110) in real time. 
The coexisting phases were characterized structurally using dark-field LEEM and $\mu$-LEED.
In dark-field LEEM, the real-space image is obtained with an aperture placed in the diffraction plane that selects a
particular LEED beam and filters out all other electrons. Thus, the dark-field image reflects the distribution of the structural domains
which gives rise to that diffraction spot. In the case that the structural domains are large enough, 
$\mu$-LEED is used to get the diffraction pattern from individual domains using an
illumination aperture to limit the incident electron beam to a micron-sized spot on the surface.

The chemical heterogeneity was mapped using i) 3p XPS core-levels of Fe and Ni in energy-filtered XPEEM mode and
ii) Fe and Ni L-edge XAS by tuning the photon energy to the respective resonance and imaging the secondary
photoelectrons. In the former, the photon energy was varied from 250~eV to 650~eV in order 
to tune the inelastic mean free path of electrons and thus to evaluate the variation of the composition 
along the surface normal. In the latter, the probing depth is several nanometers due to the large mean free path
of the secondary photoelectrons.

The W(110) substrate was cleaned with cycles of annealing in oxygen (typically 1100~$^\circ$C at $1\times10^{-6}$~mbar O$_2$
for about 15~minutes in a preparation chamber attached to the microscope) 
followed by high temperature flashes (to above 2000~$^\circ$C) in UHV to remove surface oxygen.
The cleanliness was checked with LEEM and LEED, which are very sensitive to the presence of contamination on the surface.
The Fe$_{1-x}$Ni$_{x}$ alloy film was grown on the W(110) substrate by codeposition of Fe and Ni 
from 2~mm thick high-purity rods installed in e-beam evaporators at a rate of about 0.25~ML/min.
The rate of each evaporant was determined within 5\%\ by following the completion of the respective 
pseudomorphic monolayer on W(110). Note that in the following sections the monolayer (ML) units will refer to
a layer pseudomorphic to W(110) unless otherwise stated. 
Fe monolayer completion was monitored in LEEM at about 300~$^{\circ}$C, at which
the step flow growth was easily resolved. The complete pseudomorphic Ni monolayer was obtained above 100~$^{\circ}$C.
Above a monolayer, the Ni layer transforms into $(1\times8)$ and $(1\times7)$ structures with successively higher packing
densities.~\cite{bauer1984} The onset of the $(1\times8)$ diffraction pattern in LEED was used to calibrate the Ni deposition rate. 
Upon Fe-Ni codeposition, the composition of the resulting alloy was determined also from the ratio of the integrated 3p core-level  
photoemission signals of Fe and Ni, after correcting for the difference in photoionization cross-sections and 
in the microscope transmission which depend on the electron kinetic energy.

\subsection{Ab-initio calculations}

In order to evaluate the vacancy formation energy and surface diffusion barrier,
we performed first-principles density-functional calculations as implemented in the pwscf code,~\cite{GiaBarBon09} 
which uses a plane-wave basis set and pseudopotentials. Exchange and correlation were described by 
the generalized-gradient approximation (GGA) in Perdew-Burke-Ernzerhof parametrization.~\cite{PerBurErn96}
It is known that local-density approximation overstabilizes the nonmagnetic hcp and fcc phases of Fe and 
that GGA describes well the Fe ground state.~\cite{StoBin08}
The choice of GGA gets further confirmation from the previous reports in which it is successfully applied
to calculate the vacancy formation energy in bcc Fe.~\cite{DomBec01,FuWilOrd04}
We employed Vanderbilt ultra-soft 
pseudopotentials generated using the following atomic configurations: $3d^74s^1$ for Fe and $3d^94s^1$ for Ni. 
The nonlinear-core correction to the exchange and correlation potentials was included for both Fe and Ni. 

Our plane-wave basis kinetic energy cutoff was 41~Ry for the wave-functions and 270~Ry for the charge density. 
A $7\times 7 \times1$ Monkhorst-Pack mesh was used for integrations over the Brillouin zone. 
A Gaussian broadening of 0.01~Ry was used for determination of the Fermi energy. 
The positions of the atoms in all calculations were relaxed until the residual forces on the atoms 
were smaller than $10^{-3}$ Ry/bohr. 
For the lattice parameter, we used our optimized value of bcc bulk Fe$_{0.67}$Ni$_{0.33}$, equal to 2.83\AA.

The monovacancy-formation energy in bcc $\rm Fe_{0.68}Ni_{0.32}$ was
calculated at constant volume (relaxing only the atomic positions in a supercell) as an average of 
vacancy formation energies in $\rm Fe_{0.67}Ni_{0.33}$ and $\rm Fe_{0.69}Ni_{0.31}$
in order to circumvent the minor differences in energy due to the variation in the supercell stoichiometry
when adding or substracting an atom. 
In this scheme, we started by evaluating the energy of the supercell 
containing 53 atoms, {\it i.e.} 36 Fe and 17 Ni atoms and a vacancy.
Subsequently, we performed two bulk calculations with the vacancy filled. 
In one calculation, the vacancy was filled with a Ni atom, and in the other with an Fe atom, 
thus obtaining the $\rm Fe_{36}Ni_{18}$ and $\rm Fe_{37}Ni_{17}$ configurations, respectively.
By averaging the energies for the two different reference configurations
the vacancy formation energy was obtained
for composition $\rm Fe_{36.5}Ni_{17.5}$ ($\rm Fe_{0.68}Ni_{0.32}$). 

The supercell size was chosen to be $3\times3\times3$ with 54 atoms, which had been shown to
be sufficient for convergence in $\alpha$-Fe calculations~\cite{FuWilOrd04}.  
Such a supercell is optimal for the compositions around $\rm Fe_{2/3}Ni_{1/3}$.
We constructed 3 different supercells, 
of which the first two feature random distributions of Ni sites
and the third one is  an ordered $\rm Fe_{0.67}Ni_{0.33}$ alloy. 
In the random supercells, 
the Ni positions did not follow any specific rule. In the ordered one, the Ni sites
were chosen randomly only in the first atomic layer, and the following layers were obtained by
shifting the positions along the $x$-direction from one layer to the next. 
For the vacancy formation energy, we performed a total of 14 different
calculations using three different supercells and 4 to 5 vacancy configurations for each one.

The calculation of the surface diffusion barriers was performed using a slab geometry. 
The surfaces were modeled by ``asymmetric slabs'' in which at one termination the atomic positions 
were frozen at their bulk positions, and all other positions were atomically relaxed. 
The adsorbate Ni atom was held fixed laterally at a given adsorption site 
(short and long bridge; top), while its $z$-coordinate and the atoms at the surface and below were relaxed.  
The periodic images of the slab were separated by a vacuum layer of 14~\AA. 
We used a $3\times3$ unit cell with a (110) surface orientation and the slab consisted of 5 atomic layers.

\section{Experimental results}

\begin{figure*}[t]
\begin{center}
\includegraphics{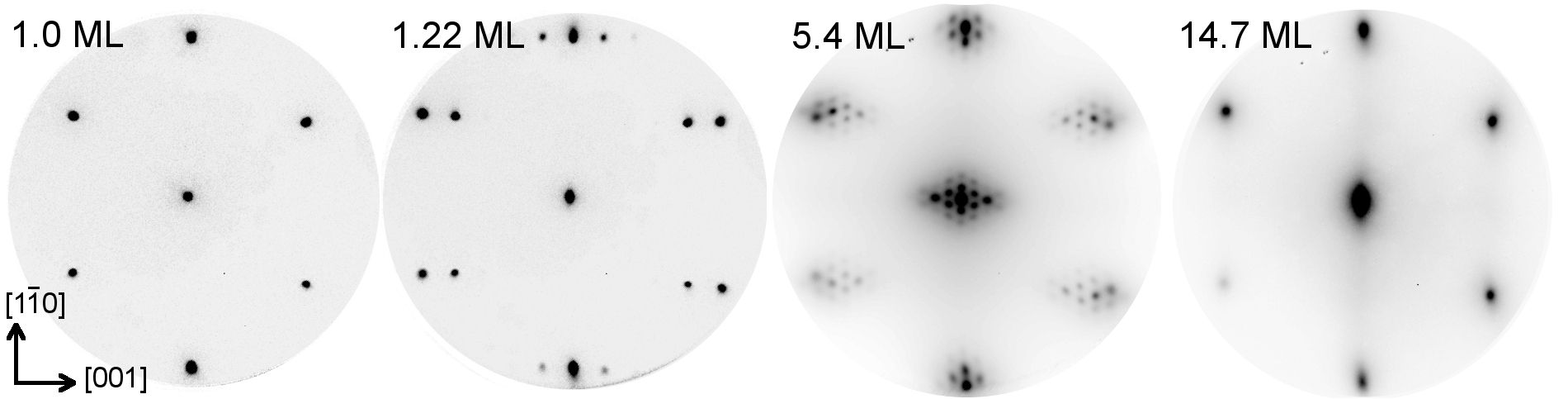}
\caption{LEED sequence on the growth of Fe$_{0.70}$Ni$_{0.30}$ on W(110) 
at room temperature by codeposition of Fe and Ni. 
The coverages are indicated in the figure. The $(1\times1)$ pseudomorphic pattern at 1.0~ML in the leftmost panel 
marks the positions of the substrate beams. The electron energy in the first two panels is 42~eV.
The last two panels, instead, are obtained by summing LEED patterns from 30~eV to 150~eV with 1~eV increments in order
to excite reflections from all diffraction orders.}
\label{fig:growth}
\end{center}
\end{figure*}

\subsection{Film growth at room temperature}

Fe$_{1-x}$Ni$_{x}$ growth by codeposition of Fe and Ni on W(110) was carried out up to
film thicknesses of 20~ML for various stoichiometries ranging from $x=0.25$ to $x=0.35$. 
For the case of $x=0.30$, the LEED data corresponding to the relevant stages during room temperature growth 
are displayed in Fig.~\ref{fig:growth}. Up to one monolayer coverage, the $(1\times1)$ substrate pattern is preserved 
without any broadening of the diffraction spots. Upon completion of the pseudomorphic 
Fe$_{0.70}$Ni$_{0.30}$ monolayer, a $(1\times8)$ superstructure appears,
the spot intensity reaching a maximum at a total FeNi coverage of 1.22~ML~\cite{mentes2015} 
as seen in the second panel in Fig.~\ref{fig:growth}. 
Interestingly, this behaviour is similar to that of a pure 
Ni layer on W(110)~\cite{bauer1984, sander1998} in spite of the Fe-rich alloy composition. 

The $(1\times7)$ structure observed for Ni/W(110) at higher coverage~\cite{stojic2012} 
is absent in the case of Fe$_{0.70}$Ni$_{0.30}$/W(110).
Above 1.22~ML, the $(1\times8)$ pattern continuously evolves into another superstructure, which fits
well with the misfit dislocation structure observed for pure Fe films on W(110).~\cite{gradmann1982}
The misfit dislocation pattern (third panel in Fig.~\ref{fig:growth}) persists up to about 10~ML. Further growth at
room temperature results in the ($1\times1$) bcc(110) pattern seen in the rightmost panel in Fig.~\ref{fig:growth}.
The position of the fractional spots indicates 
approximately 10\%\ contraction with respect to the W(110) lattice, in good agreement with the Fe(110)
surface. Moreover, the broadening of the diffraction spots
and the streaks along the $[1\bar{1}0]$ direction are also
observed for Fe films on W(110) of comparable thickness. 
The streaks contain weak diffuse facet spots that move in reciprocal space along $[1\bar{1}0]$ as a function of energy
(a direct consequence of faceting),
as in the case of Fe/W(110).

The surface of films prepared at room temperature has a slight granularity below the 50~nm length scale.
However, the presence of fcc domains is ruled out based on the LEED and dark-field LEEM data. 
Moreover, XPEEM imaging at the Fe 3p and Ni 3p core levels show no chemical heterogeneity above 
the resolution limit of about 30~nm. Possibly, this roughness is associated with the nanoscale faceting of the room
temperature grown surface as hinted in the elongated spot profiles in the rightmost panel in Fig.~\ref{fig:growth}. 
As we will discuss in the following sections, the fcc-bcc phase separation 
upon annealing takes place at a much longer length scale than the disorder present in the room temperature grown films.

\subsection{Nucleation and growth of {\it fcc} regions at high temperature}

As illustrated in Fig.~\ref{fig:growth}, a room temperature codeposition of Fe and Ni 
up to a total thickness above 10~ML results in bcc(110) Fe$_{0.70}$Ni$_{0.30}$ films. 
The disorder in the room-temperature grown films, visible in the broad diffraction beam profiles,
smoothens out upon annealing the surface to temperatures below 250$^{\circ}$C without
changing the bcc order. At about 400$^{\circ}$C and above, the film breaks apart. However, annealing to
temperatures between 250$^{\circ}$C to 400$^{\circ}$C
an additional LEED pattern of hexagonal symmetry appears, accompanied
by the formation of islands at the micron scale. 

\begin{figure}[b]
\begin{center}
\includegraphics[width=8.0cm]{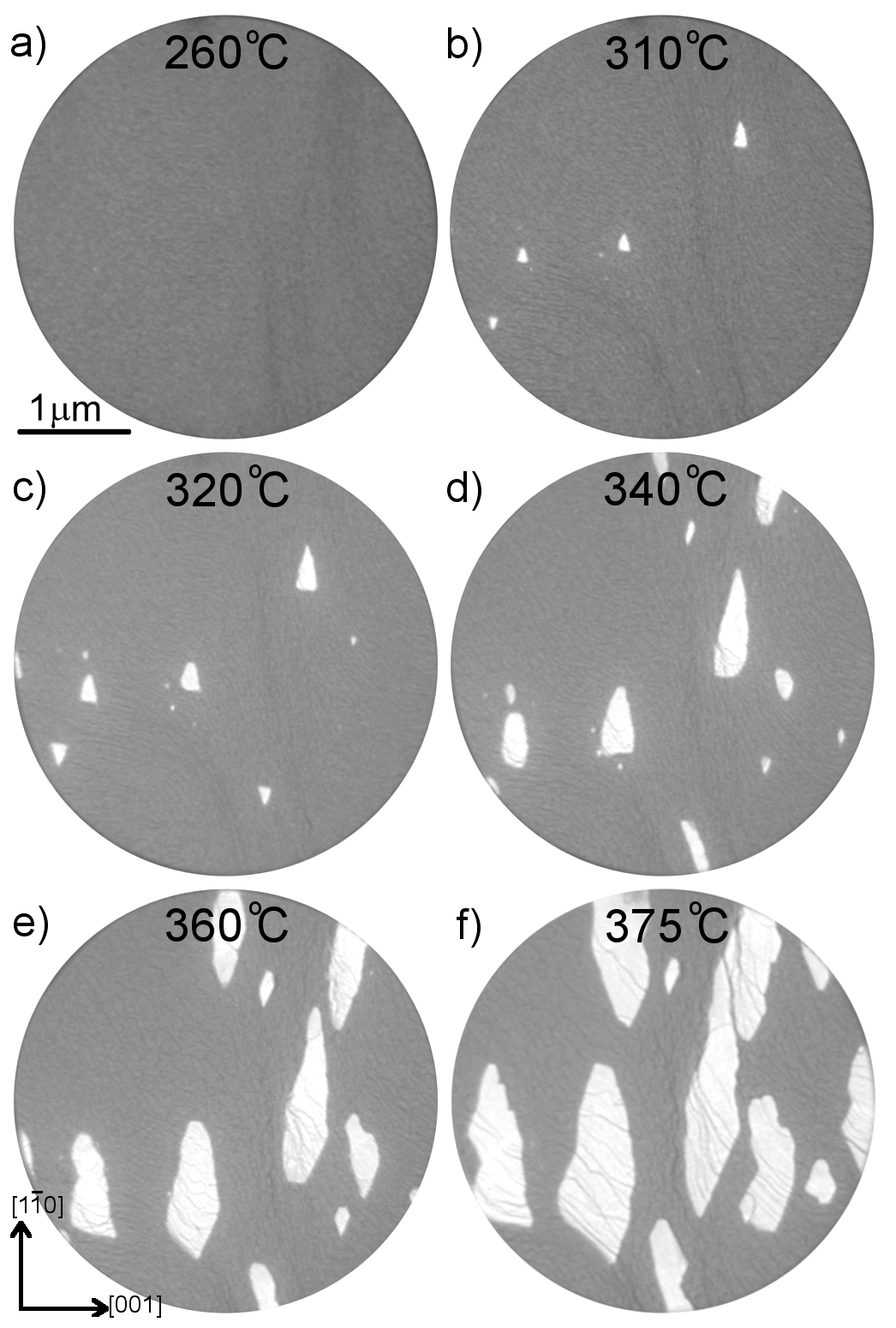}
\caption{Annealing of 18~ML Fe$_{0.70}$Ni$_{0.30}$ film in LEEM.
a) 260$^{\circ}$C, b) 310$^{\circ}$C, c) 320$^{\circ}$C, d) 340$^{\circ}$C, e) 360$^{\circ}$C, f) 375$^{\circ}$C.
Electron energy is 19~eV. The temperature ramp from (a) to (f) spans a period of 1900~s.}
\label{fig:anneal}
\end{center}
\end{figure}

The growth of fcc regions for an 18~ML Fe$_{0.70}$Ni$_{0.30}$ film 
can be followed in the LEEM sequence shown in Fig.~\ref{fig:anneal}.
Up to above 260$^{\circ}$C the surface remains homogeneous with a uniform change in electron reflectivity due to a
progressive flattening/ordering of the surface. Following a slow temperature ramp (about 10$^{\circ}$C/min), 
triangular regions appear at about 300$^{\circ}$C. These grow slowly at low temperature and faster at higher temperature.
Experiments for different film stoichiometries indicate that
alloys with higher Ni content tend towards a larger
surface coverage of islands. The island shapes are elongated in the bcc $[1\bar{1}0]$ direction.
This is qualitatively consistent with the faster diffusion along $[1\bar{1}0]$. However, the measured
aspect ratio of the islands is near 2, which describes a more anisotropic shape than what is expected from surface
diffusion alone. We tentatively attribute the island shape to (at least in part) the boundary energy between the
fcc and bcc domains.

\begin{figure}
\includegraphics{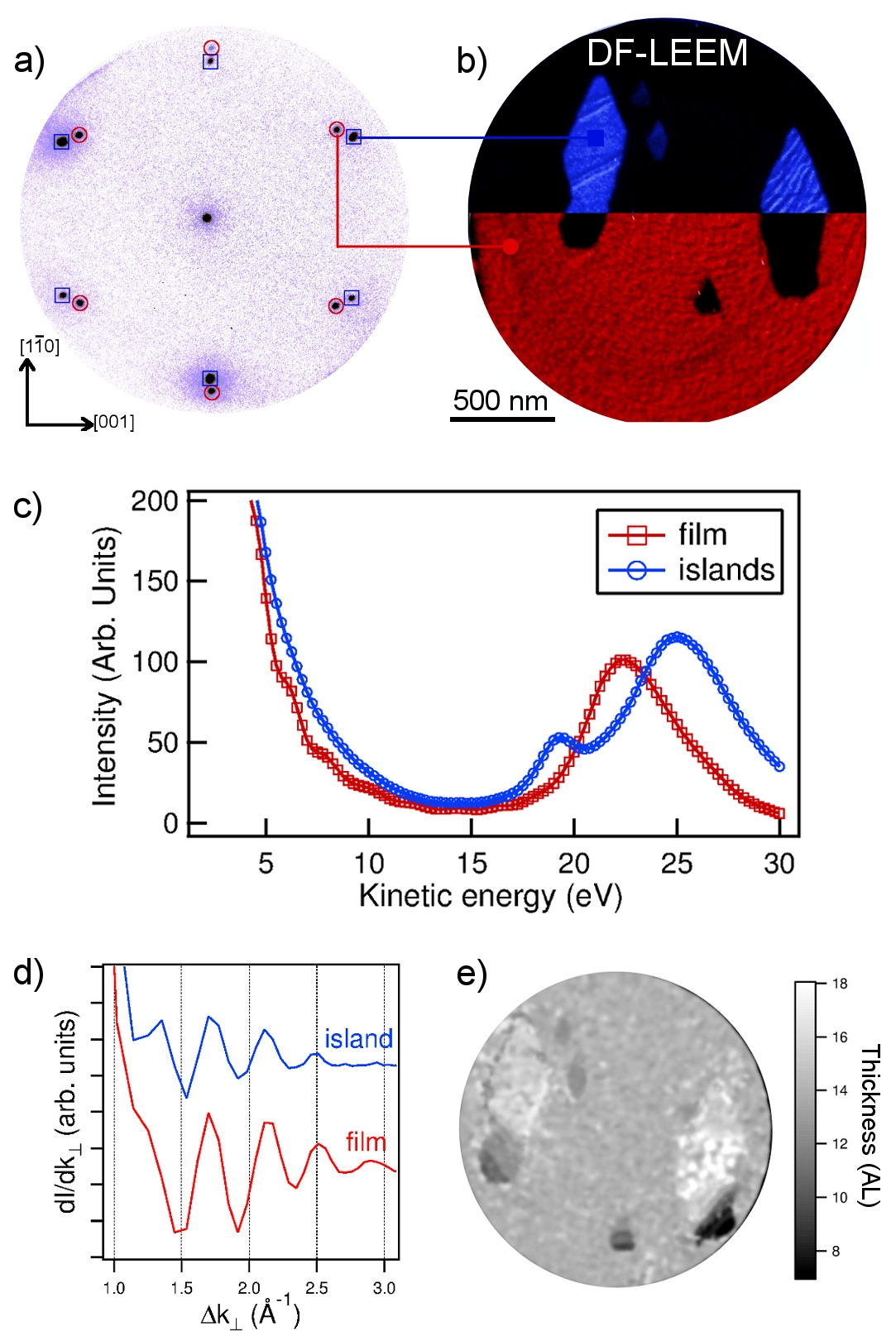}
\caption{(Color online) a) LEED pattern at 54~eV. The bcc pattern is marked by (red) circles, and the hexagonal structure
is marked by (blue) squares. b) Combined dark field LEEM 
images. The upper half is obtained by imaging with the fcc spot, whereas the lower half using the bcc spot. Electron energy is 60~eV.
c) LEEM I(V) curves from the two regions above.
d) Differential LEEM I(V) data showing the quantum oscillations in electron reflectivity.
e) Film thickness map derived from the oscillation period in the LEEM I(V) data.
Thickness is indicated in atomic layers (AL), which is about 20\%\ denser than the pseudomorphic ML.}
\label{fig:characterize}
\end{figure}

The LEED pattern of such a heterogeneous surface is the superposition of two ($1\times1$) unit cells 
as shown in Fig.~\ref{fig:characterize}a. One of the two corresponds to the symmetry of a bcc(110) surface, with the
lattice constant very close to that of a Fe(110) surface. Using the W substrate lattice as a reference, the lattice constant of
this bcc pattern is very slightly smaller (by 0.5\%) than that of Fe(110).
The other LEED pattern represents a hexagonal structure and has a lattice vector expanded by 2.2\%\ 
with respect to the Ni fcc(111) unit cell.

Dark-field LEEM images (Fig.~\ref{fig:characterize}b)
indicate that the hexagonal pattern originates from the islands whereas the rest of the film has the bcc(110) structure.
In order to better understand the structure of the islands, the energy dependent electron reflectivity curves for the two regions 
are plotted in Fig.~\ref{fig:characterize}c. The islands show a characteristic double peak I(V) spectrum at around the
first Bragg energy, which is very similar to the I(V) curve of the clean Ni(111) surface.~\cite{flege2011}
Therefore, based on the hexagonal kinematic LEED pattern and the energy dependence of electron reflectivity,
we conclude that the islands have the fcc structure.

A careful inspection of the LEEM I(V) data reveals that thickness-dependent intensity oscillations are not uniform
over the surface. Fig.~\ref{fig:characterize}d displays the intensity derivative versus out-of-plane momentum transfer, in
which the regular quantum oscillations are evident. The film thickness is directly related to these oscillations,~\cite{altman2010}
with thicker films resulting in a shorter period and a smaller amplitude 
due to reflectivity attenuation effects induced by the limited electron mean free path.
Thus, the thickness can be laterally mapped from the local I(V) spectra, as shown in Fig.~\ref{fig:characterize}e. 
The large fcc islands are thicker than the bcc regions by about 15\%\ on average, 
although they also feature regions with reduced thickness. The smaller islands
instead are systematically thinner than the rest of the film. Importantly, the variations in the thickness point to a mechanism
involving mass transport over distances above a micron. Moreover, the well-defined quantum oscillations in electron reflectivity
show that the structure is homogeneous along the entire film depth both in the bcc and fcc regions.

\begin{figure*}
\includegraphics{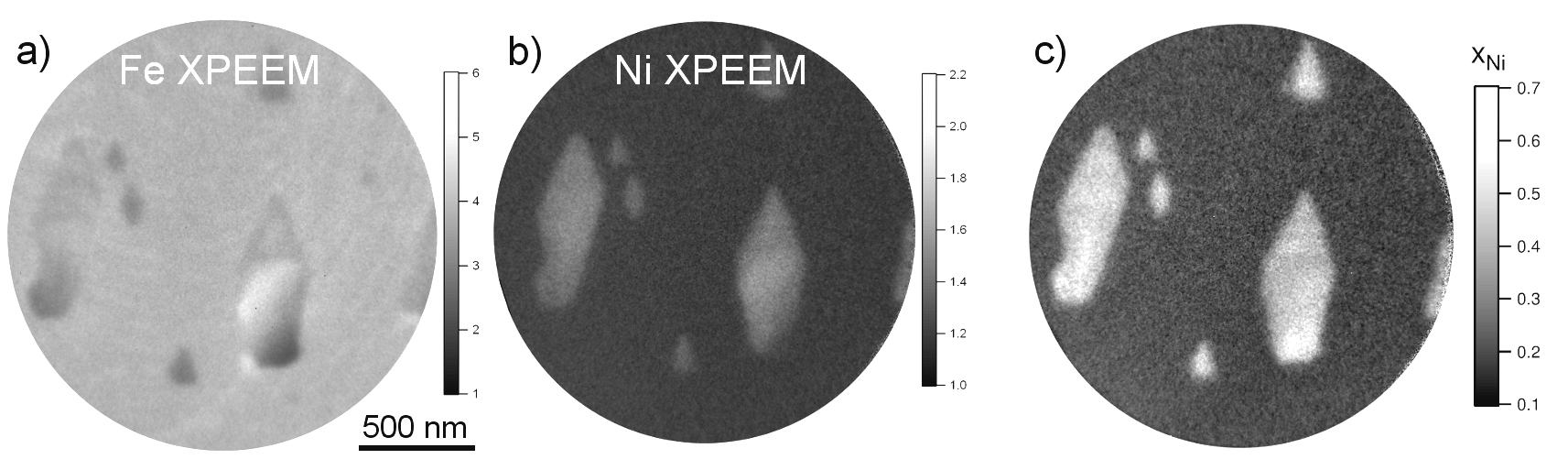}
\caption{a,b) XPEEM images acquired at the Fe (a) and Ni (b) L$_3$ absorption edges at about 
708~eV and 853~eV, respectively. The images, which are normalized to those before the absorption threshold, 
reflect the XAS white line intensity given in the gray scale bars for each image. c) The resulting stoichiometry map
showing the Ni content.}
\label{fig:XPEEM}
\end{figure*}

The question arises to the chemical composition of the structurally heterogeneous surface as revealed by LEEM.
Laterally resolved Fe and Ni distributions are displayed in the XAS-PEEM images in Fig.~\ref{fig:XPEEM}a and b, respectively.
The chemical maps are given by the secondary photoelectron images obtained at photon energies around 
the L$_3$ x-ray absorption threshold of the respective element.
More precisely, the image acquired with a photon energy at
the center of resonance (resonant image) is normalized to that acquired at a photon energy several eV below 
the resonance (baseline image). In such a normalized XAS-PEEM image, a value of unity corresponds to the absence of the
respective element. 
Using the average composition ($x=0.30$) and the area coverage of the two phases as input, 
the lateral map of the Fe$_{1-x}$Ni$_{x}$ alloy stoichiometry ($x$) can be obtained from the x-ray intensity images,
as seen in Fig.~\ref{fig:XPEEM}c. Qualitatively, Fig.~\ref{fig:XPEEM}c shows that i) there is a clear composition difference
between the two phases, and ii) the composition is nearly independent of the thickness variations.
Quantitatively, we can assign $x = 0.23 \pm 0.03$ and $x = 0.49 \pm 0.05$ to the bcc and fcc phases, respectively.
The pronounced variation in the composition is a direct proof that the bcc-fcc phase separation is not due to a local or
martensitic structural transformation but instead it involves considerable mass transport.

Beyond determining the compositions, the XPEEM images 
show a qualitative difference between Fe and Ni distributions.
Apart from the variations correlated to film thickness, 
the Fe signal is rather uniform everywhere on the surface. Instead, the fcc islands have more than 
a twice stronger Ni signal compared to the bcc film. 
We note that the XAS data displayed in Figs.~\ref{fig:XPEEM}a and \ref{fig:XPEEM}b reflect
the total Fe and Ni amounts, respectively, along the film normal 
due to the relatively long mean free path of the secondary photoelectrons. Therefore,
these observations suggest that the chemical heterogeneity is formed 
by Ni mass transport laterally across the film, whereas Fe mostly remains in place.
The fact that the fcc islands are thicker than the nominal film thickness confirm this statement.
However, the quantitative difference in thickness between the fcc and bcc regions is not sufficient to rule out lateral Fe diffusion.

We have also acquired x-ray photoemission spectra at the Fe 3p and Ni 3p core levels for a different
nominal alloy composition (shown in the supplementary material). Importantly, the compositions extracted from the
laterally-resolved XPS data are dependent on the photon energy. In particular, both phases appear more Ni rich at
energies with shorter electron inelastic mean free path, indicating that the surface is enriched in Ni.

Finally, we note that the fcc-bcc coexistence was observed in the entire composition range under study 
($x=0.25$ to $0.35$). Qualitatively, the same behaviour was observed in all experiments. Importantly, 
at the highest Ni content considered ($x=0.35$) the fcc structure was present already 
after the room temperature growth and before annealing. Moreover,
the length scale at which the fcc-bcc separation takes place was observed to be considerably shorter.
At compositions below $x=0.25$, the bcc phase was stable up to the dewetting 
temperature of the film, which is at around 400$^{\circ}$C for the thicknesses considered.

\subsection{Phase separation kinetics}

\begin{figure}
\includegraphics[width=8.0cm]{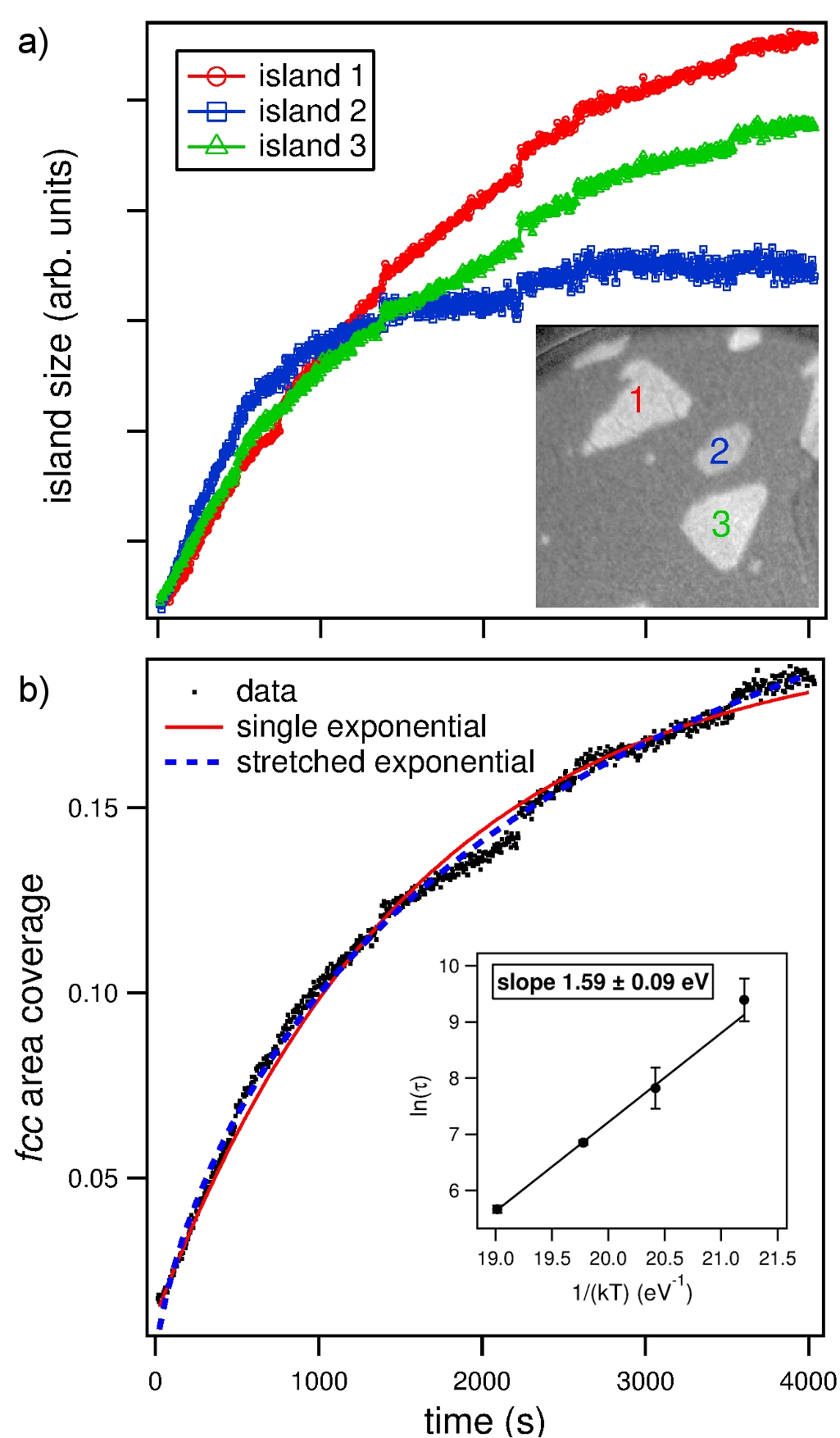}
\caption{(Color online) Growth of fcc islands followed in LEEM. Electron energy is 19~eV. The increase in (a) the area of individual islands
labeled in the inset as a funtion of time, (b) overall fcc island coverage as a function of time. At  $t=0\ s$, 
the temperature is increased from below 200$^{\circ}$C to 340$^{\circ}$C and is kept constant afterwards. 
The inset shows the time constant of the exponential fit vs inverse temperature
obtained from a series of similar experiments at varying temperatures.
The small abrupt variations in the measured island coverage is due to readjustments of the objective focus, which
slowly drifts at high temperature.}
\label{fig:kinetics}
\end{figure}

The fcc island growth was monitored in real time using LEEM. The result is given in Fig.~\ref{fig:kinetics}. Firstly, we observe that
the island coverage tends to a certain value for the given nominal composition, and that there is a variation in the growth of
individual islands. The latter is evidenced in Fig.~\ref{fig:kinetics}a, in which the growth of the fcc island labeled `2' seems
impeded by the growth of the larger islands nearby. This is consistent with material exchange over long distances, which
causes a competition between nearby islands in capturing material.

Fig.~\ref{fig:kinetics}b shows a plot of the fractional fcc island coverage as a function of time at constant temperature
(340$^{\circ}$C). Notably, the time dependence is reproduced reasonably well with a single exponential function. Similar experiments
at different temperatures show that the time constant has an Arrhenius behaviour. In other words, the temperature-dependence
of the dynamics suggests the presence of an energy barrier of $1.59\pm0.09$~eV. The final fcc coverage is 25\%\ for the
data displayed in Fig.~\ref{fig:kinetics}. Other measurements qualitatively indicate that this value depends on the nominal
composition, with higher Ni content favoring the abundance of the fcc structure.

Although the single exponential function approximately represents the entire time scale, 
the data in Fig.~\ref{fig:kinetics}b show a faster rise immediately 
after nucleation and a longer tail at later stages of island growth.
This is an indication that there is more than one energy barrier contributing to the relaxation towards equilibrium. 
Therefore, we consider the more general case, in which there is a distribution of barriers. The excellent fit in 
Fig.~\ref{fig:kinetics}b shows that the time evolution can be represented by a stretched exponential function,~\cite{santos2005}
which is often applied to relaxation in disordered systems. 
The stretched exponential is described by a time dependence in the form $e^{-(t/\tau)^\beta}$.
The exponent $\beta$ is found to be 0.70 for the data displayed in Fig.~\ref{fig:kinetics}b.

The unitless quantity $\beta$ carries information on the distribution of energy barriers.~\cite{edholm2000} 
In particular, the width of the barrier distribution scales as $k_{B}T/\beta$. Taking a Gaussian barrier distribution, the full width at half
maximum (FWHM) is found to be about 0.17~eV for $\beta=0.70$. 
In order to give a physical sense to this energy barrier distribution, in the following section we will present 
the results of DFT calculations for the energy parameters associated with the bcc $\rm Fe_{2/3}Ni_{1/3}$ alloy.

\section{Theoretical calculations}

\begin{figure*}
\includegraphics{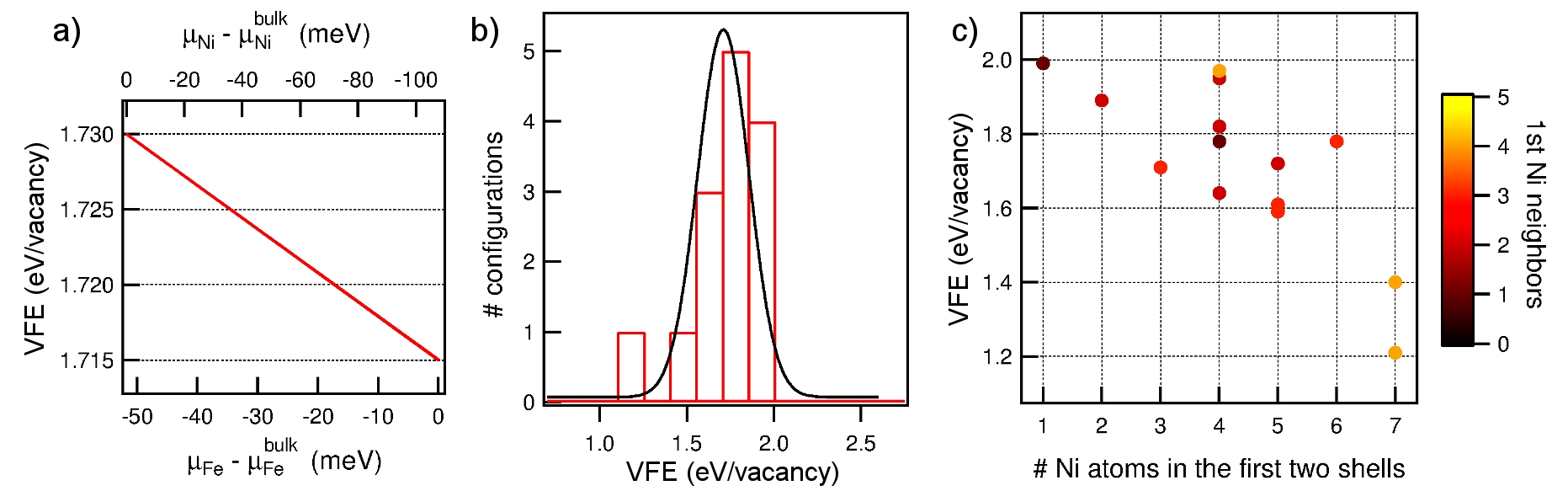}
\caption{(Color online) a) One example of vacancy formation energy (VFE) of the ordered bulk Fe$_{0.68}$Ni$_{0.32}$ 
performed for a vacancy configuration including 3 first and no second Ni neighbors. The lower horizontal 
axis represents the difference between the Fe chemical potential in the alloy  Fe$_{0.68}$Ni$_{0.32}$, $\mu_{Fe}$, 
and in the elemental bulk phase, $\mu_{Fe}^{bulk}$. 
The corresponding difference for Ni is shown in the upper horizontal axis. 
b) Histogram of the vacancy formation energies obtained from 14 calculations. 
c) The vacancy formation energies vs the total number of Ni atoms in the first two shells around the vacancy.  
The color coding denotes the number of first Ni neighbors. }  
\label{fig:vac_en}
\end{figure*}

In Figure~\ref{fig:vac_en}a we show an example of the calculated vacancy formation energy (VFE)
for the Fe$_{0.68}$Ni$_{0.32}$ alloy. It is given as a function of  the chemical potential
difference between the alloy and the elemental bulk phases, 
shown for Fe and Ni respectively on the bottom and top horizontal axes.
The maximum range of the chemical potential differences are derived from the calculated formation enthalpy,
which is $-0.035$~eV/atom for the particular supercell in Figure~\ref{fig:vac_en}a.
As can be seen, the vacancy formation energy depends only weakly
on the chemical potential. 
Moreover, this dependence is identical for all the calculations performed (see supplementary material).
Therefore, for simplicity, in the following we will use the values of vacancy formation energies 
for the Fe chemical potential equal to the bulk Fe chemical potential. 

In order to account for a possible influence of the immediate environment of the vacancy, we have
performed calculations keeping track of the number of Ni atoms near the vacancy. 
In particular, the stoichiometry and the randomness of the vacancy and Ni positions are taken into account
in dividing the total of 14 calculations according to the number of first Ni neighbours
into groups of 5 calculated configurations with 2 first Ni neighbors, 
4 with 3, 3 with 4, and 2 with a single first Ni neighbor. 
This choice roughly corresponds to the weights~\cite{delSamDog11} 
each number of neighbors has in a random Fe$_{0.68}$Ni$_{0.32}$ alloy.
Note that we have not considered vacancy configurations with 5 to 8 Ni neighbours, as they have a negligible
contribution at this stoichiometry.

Figure~\ref{fig:vac_en}b presents a histogram of the resulting vacancy formation energies.
The histogram can be roughly described by a Gaussian distribution 
centered around 1.71~eV with a FWHM deviation of 0.37~eV. These values
are in good agreement with the very recent results obtained using kinetic Monte Carlo calculations, which reported
a range of vacancy formation energies between 1.2~eV and 2.15~eV for Fe$_{0.50}$Ni$_{0.50}$.~\cite{belsamsto2016}

Figure~\ref{fig:vac_en}c shows the vacancy formation energy 
as a function of the total number of Ni neighbors in the first two shells.  
From the clear correlation,
it can be inferred that the vacancy formation energy is reduced
for configurations with increasing number of Ni neighbors. 
This is in line with our observation that the Ni atoms, constrained in the bcc environment, 
relax significantly more than the Fe atoms around the vacancy, 
and, in consequence, tend to reduce the vacancy formation energy. 
We note that the strongest Ni relaxations occur when more Ni atoms are grouped together rather than
sitting on scattered sites around the vacancy.

The tendency towards a value somewhat above 2~eV/vacancy in the absence of Ni neighbours can be seen
in Figure~\ref{fig:vac_en}c. This is
consistent with the theoretical vacancy formation energy of 2.2~eV in pure Fe.~\cite{FuWilOrd04,chobartuc2011}
Nevertheless, we also encounter differences in the vacancy energy for equivalent configurations of Ni neighbours
near the vacancy within different supercells. We attribute such differences to long range elastic interactions, which
depend on the entire supercell configuration.

Lastly, we also calculate the diffusion barrier for a Ni adatom on the bcc alloy surface. 
We focus on the surface diffusion of Ni adatoms because of
the experimental observation that the surface is enriched in Ni, as reported in the previous section.
The lowest energy path is found to be along the direction connecting the long-bridge site (the global energy minimum)
and the short-bridge site, i.e. along the $[1\bar{1}1]$ direction. The diffusion barrier is calculated to be 0.34~eV.
This value compares well with the calculation of a Fe/Fe(110) diffusion barrier of 0.36~eV,~\cite{chegiogui2012}
and is considerably lower than the barrier of 0.65~eV for Ni/Ni(111).~\cite{ondrajswi2006}

\section{Discussion}

The experimental results reported in the preceding sections showed that there is a structural and chemical
rearrangement at the micron scale of the Fe$_{1-x}$Ni$_{x}$ alloy film close to $x=0.30$. 
Vacancy formation and surface diffusion are the two important ingredients in the restructuring of the film.
It has been known that the surface of Fe-Ni alloys has a tendency 
to be enriched in Ni.~\cite{wandelt1976,sanribdho2016} By varying the probing depth via the inelastic
mean free path of photoelectrons in photon energy dependent XPS (in addition to the XAS spectra acquired with the
secondary photoelectrons) we found that
(i) there is no surface segregation in the room-temperature grown films, 
(ii) upon annealing, the surface is enriched in Ni both in the fcc and bcc phases.
In short, we attribute the chemical heterogeneity predominantly to the surface diffusion of Ni, with the
calculated diffusion barrier of 0.34~eV as reported in the previous section.

The calculated average vacancy formation energy of 1.71~eV/vacancy is in good agreement with the experimental energy barrier of 
$1.59 \pm 0.09$~eV found from the phase separation kinetics. Moreover, the experimental kinetics indicates the presence of
a barrier distribution with a FWHM of 0.17~eV. This is also reproduced in the calculated distribution of vacancy formation
energies, although with a larger spread of 0.37~eV. Therefore, considering the much lower surface diffusion barrier,
we conclude that the rate limiting step in the phase separation in ultrathin alloy films is vacancy formation.

The difference between the theoretical vacancy formation energy and the experimentally evaluated barrier lies
in the temperature dependence of the vacancy formation {\em free} energy. It has been recently reported that there is a pronounced
change in the energy parameters associated with mono-vacancies in bcc Fe as a function of temperature.~\cite{wenwoo2016}
In particular, the reported decrease in the vacancy formation free energy from 0~K to 600~K is about 10\%\, which
quantitatively agrees with the difference between our zero-temperature DFT calculations and 
our experimental result at 600~K for the bcc Fe-Ni alloy.

As opposed to time dependence, the length scale at which the phase separation takes place reflects
the pronounced surface diffusion at around 300$^{\circ}$C. Moreover, the fcc island shapes are elongated along 
the $[1\bar{1}0]$ direction, which is the faster diffusion direction on the bcc(110) surface. Therefore, we attribute 
the mesoscopic morphology of the phase-separated film to surface diffusion, while the entire phase-separation
and the atomic restructuring process is limited by vacancy formation.

Importantly, these results clarify the main distinction between an alloy film (or the alloy surface) and a bulk Fe-Ni alloy.
Although the surface appears crucial in determining the microscale morphology, interstitial diffusion barriers in the bcc Fe 
and its alloys are just as low as the surface diffusion barriers.~\cite{FuWilOrd04,chobartuc2011,marwilmou2011}
Instead, the interstitial formation energy is considerably larger than that of vacancies, and our study singles out vacancy formation
as the limiting step in the phase separation process. Thus, the principal role of the surface in facilitating the phase separation is the
creation of vacancies without forming interstitials.

\section{Conclusion}

In summary, we have experimentally shown that the Fe$_{0.70}$Ni$_{0.30}$/W(110) film
of thickness close to 20 atomic layers separates into micron-sized fcc and bcc
regions at around 300$^{\circ}$C. The two structural regions have distinctly different compositions and the composition
difference originates from a difference in the Ni amount. In addition, the energy barrier associated with the kinetics 
of phase separation matches the theoretically calculated vacancy formation energy. Moreover, the calculated vacancy
formation energy is slightly different for each vacancy configuration, giving a distribution of energy barriers in agreement
with experimental observation. In the phase separation process, the surface facilitates the creation of vacancies, 
as well as helping Ni diffusion with surface Ni enrichment and the pronounced diffusion length.
We expect that the observed fcc-bcc phase separation in Fe-Ni thin films at the mesoscopic scale
will provide a model for studying the effect of chemical and structural heterogeneity on the alloy magnetic properties 
near the Invar composition.

\begin{acknowledgments}
We gratefully acknowledge Nadia Binggeli and Mighfar Imam for useful discussions.
\end{acknowledgments}


\end{document}